\documentclass[journal]{IEEEtran} % use the `journal` option for ITherm conference style
\IEEEoverridecommandlockouts
\usepackage{cite}
\usepackage{amsmath,amssymb,amsfonts}
\usepackage{algorithmic}
\usepackage{graphicx}
\usepackage{textcomp}
\usepackage{xcolor}
\usepackage{color,soul}
\usepackage{amsbsy}
\usepackage{amsfonts}
\usepackage{mathrsfs}
\usepackage{amsmath} % assumes amsmath package installed
\usepackage{amssymb}  % assumes amsmath package installed
\usepackage{lineno}
\usepackage{epstopdf}
\usepackage{graphicx}
\usepackage{setspace}
\usepackage{booktabs}
\usepackage{stfloats}
\usepackage{indentfirst}
\usepackage{caption}
\usepackage{cite}
\usepackage[T1]{fontenc} % optional
\usepackage{bm} % optional
\usepackage{framed}
\usepackage{multicol}
\usepackage{cuted}
\usepackage{subfigure}
\usepackage{threeparttable}
\usepackage{enumerate,multirow}
\usepackage[mathscr]{eucal}
\usepackage{times}
\usepackage{amstext}
\usepackage{stmaryrd}
\usepackage{graphicx}
\usepackage{epstopdf}
\usepackage{array}
\usepackage{longtable}
\usepackage{multirow}
\usepackage[figresright]{rotating}
\usepackage{amsmath}
\usepackage{epsfig}
\usepackage[caption=false, font=footnotesize]{subfig}
\usepackage{nomencl}
\usepackage{booktabs}
\usepackage{algorithm}

\def\BibTeX{{\rm B\kern-.05em{\sc i\kern-.025em b}\kern-.08em
    T\kern-.1667em\lower.7ex\hbox{E}\kern-.125emX}}

\newtheorem{lemma}{Lemma}

\begin{document}

\title{A Tutorial on Linear Least Square Estimation\\
% delete or comment-out the following line before submission
}
%{\footnotesize \textsuperscript{*}Note: Sub-titles are not captured in Xplore and should not be used}
%\thanks{Identify applicable funding agency here. If none, delete this.}

\author{%%%% author names
    \IEEEauthorblockN{ Qingrui Zhang}% first author
    % duplicate the line above as many times as needed to list all authors
    \\%%%% author affiliations
    \IEEEauthorblockA{\textit{Sun Yat-sen University,  China}}\\% first affiliation
    % duplicate the line above as many times as needed to list all affiliations
    %%%% corresponding author contact details
    \IEEEauthorblockA{zhangqr9@mail.sysu.edu.cn}
}

\maketitle

\begin{abstract}
    This is a brief tutorial on the least square estimation technique that is straightforward yet effective for parameter estimation. The tutorial is focused on the linear LSE's instead of nonlinear versions, since most nonlinear LSE's can be approximated non-trivially using its linear counterparts. Linear LSE's can also provide insight into the study of the nonlinear techniques, e.g.  Gauss-Newton method and Lavenberg–Marquardt method etc. Linear LSE's are computationally efficient for most occasions, so they are widely applied in practice.   In this tutorial, both the original batch least square estimation and its recursive variants are reviewed comprehensively with detailed mathematical derivations. 
\end{abstract}


%%-------8-------8-------8-------8-------8------- New Section -------8-------8-------8-------8-------8-------8-------
\section{Introduction} \label{sec:Intro}
Least square approach is one of the most popular estimation techniques that has been widely applied to diverse fields such
as machine learning, system identification, and adaptive control etc. The least square estimation (LSE) is mainly used to approximate a parameter of a model by measuring the system inputs and outputs, so it also called ``linear regression'' in statistics. In LSE, the model parameter is approximated by minimizing the squared difference between the observed data and their expected values, also called ``mean squared error''.  The linear LSE is, therefore, formulated as a convex optimization problem whose solution could be found by setting the gradient to be zero.  

LSE is commonly used in the case where we expect a linear relationship between measured variables. For instance, the applied force $F$ on a spring system has a linear relationship with the spring displacement $y$, namely $F=ky$, according to Hooke's law. The spring constant $k$ hereby is unknown, so it could be approximated using LSE with a batch of measurements on $F$ and $y$. Those measurements are also called data samples. Another example is the relationship between the angle of attack $\alpha$ and the lift $L$ on the wing of an aircraft. In a flight with a constant velocity $V$, the relationship could be simplified into $L = \frac{1}{2}\rho V^2 S \left(C_{L0}+C_{L\alpha}\alpha\right)$, where $\rho$ is the air density and $S$ is the wing area \cite{Beard2012Book,Zhang2019PhD}. Both  $\rho$ and $S$ are known constant parameters. Hence, the approximation of the aerodynamic coefficients $C_{L0}$ and $C_{L\alpha}$ could be estimated based on the LSE technique with a set of measurements on $L$ and $\alpha$.

In this tutorial, the standard LSE problem will be formulated firstly. It follows by the original batch LSE design for constant parameter estimation. After the batch LSE, a recursive variant will be presented. The recursive LSE is computationally more efficient than its batch counterpart, so it could be applied in a real-time fashion. Afterwards, the recursive LSE with forgetting factor is also reviewed, which is potentially used to estimate time-varying parameters.

%%-------8-------8-------8-------8-------8------- New Section -------8-------8-------8-------8-------8-------8-------
\section{Preliminaries} \label{sec:Prelim}
In this section, some preliminaries will be provided, which will be used in the derivation of the recursive LSE \cite{Tylavsky1981Proceed}. 
\begin{lemma}[Woodbury matrix identity] \label{lem:Woodbury}
Suppose $\boldsymbol{A}\in\mathbb{R}^{n\times n}$ and $\boldsymbol{C}\in\mathbb{R}^{m\times m}$ are both square and invertible matrices, $U\in\mathbb{R}^{n\times m}$ and $V\in\mathbb{R}^{m\times n}$ are appropriate matrices, then the following identity holds.
\begin{align}
\left(\boldsymbol{A}+\boldsymbol{U}\boldsymbol{C}\boldsymbol{V}\right)^{-1}=&\boldsymbol{A}^{-1}-\boldsymbol{A}^ {-1}\boldsymbol{U}\left(\boldsymbol{C}^{-1}+\boldsymbol{V}\boldsymbol{A}^{-1}\boldsymbol{U}\right)^{-1} \nonumber\\
&\times\boldsymbol{VA}^{-1} \label{eq: Woodbury_MI}
\end{align}
\end{lemma}
\begin{lemma}[Sherman-Morrison
formula] \label{lem: Sherman-Morrison}
Consider an invertible square matrix $\boldsymbol{A}\in\mathbb{R}^{n\times n}$ and column matrices $\boldsymbol{u}$, $\boldsymbol{v}\in\mathbb{R}^{n\times 1}$, the following relationship exists.
\begin{equation}
\left(\boldsymbol{A}+\boldsymbol{u}\boldsymbol{v}^T\right)^{-1}=\boldsymbol{A}^{-1}-\frac{\boldsymbol{A}^ {-1}\boldsymbol{u}\boldsymbol{v}^T\boldsymbol{A}^{-1}}{1+\boldsymbol{v}^T\boldsymbol{A}^{-1}\boldsymbol{u}}\label{eq: Sherman-Morrison}
\end{equation}
\end{lemma}
Note that Lemma \ref{lem:Woodbury} is also named by the matrix inversion lemma in some references. Lemma \ref{lem: Sherman-Morrison} is a special case of the Woodbury matrix identity in Lemma \ref{lem:Woodbury}.

%%-------8-------8-------8-------8-------8------- New Section -------8-------8-------8-------8-------8-------8-------
\section{Problem formulation} \label{sec:ProbForm}
In linear LSE,  a system output or observation denoted by $\boldsymbol{Y}$ has a linear relationship with system inputs or states specified by $\boldsymbol{X}_i$ as given below.
\begin{align}
\boldsymbol{Y} =& \theta_0 + \theta_1  \boldsymbol{X}_1+\ldots+ \theta_n  \boldsymbol{X}_n \label{eq: LSE_Prob}
\end{align}
where $\theta_0$, $\theta_1$, $\ldots$, $\theta_n$ are unknown parameters. In the implementation, it is assumed that $\boldsymbol{Y}$ and $\boldsymbol{X}_i$ can be measured or observed. The objective is to find a set of parameters $\hat{\theta}_0$, $\hat{\theta}_1$, $\ldots$, $\hat{\theta}_n$ as an estimates of $\theta_0$, $\theta_1$, $\ldots$, $\theta_n$, respectively, such that the following function is minimized.
\begin{equation}
\min_{\theta_0,\theta_1,\ldots,\theta_n} \Vert \hat{\boldsymbol{Y}}- \boldsymbol{Y}\Vert_2
\end{equation}
where $\boldsymbol{Y}$ is observed outputs and $ \hat{\boldsymbol{Y}} = \hat{\theta}_0+\hat{\theta}_1 \boldsymbol{X}_1+\ldots+ \hat{\theta}_n\boldsymbol{X}_n$.

%%-------8-------8-------8-------8-------8------- New Section -------8-------8-------8-------8-------8-------8-------
\section{Batch least square estimation}  \label{sec:BatchLSE}
According to Section \ref{sec:ProbForm}, consider the following model with a scalar output and $n$ inputs.
\begin{equation}
y\left(k\right) = \sum_{i=1}^nx_i\left(k-1\right)\theta_i=\boldsymbol{\phi}\left(k-1\right)^T\boldsymbol{\theta}
\end{equation}
where 
$\boldsymbol{\phi}\left(k\right)=\left[x_0\left(k\right),x_1\left(k\right),\ldots,x_n\left(k\right)\right]^T\in\mathbb{R}^{n\times 1}$
is a \textbf{known} and \textbf{measurable} function,
$y\left(k\right)\in\mathbb{R}$ is the observed ouput, $k$ is the time step, and
$\boldsymbol{\theta}=\left[\theta_0,\theta_1,\ldots,\theta_n\right]^T\in\mathbb{R}^{n\times 1}$
is the \textbf{unknown} but constant parameter to be estimated.

Assume the system is observed for up to $k$ times with the following
data are collected.
\[y(1)\text{, }y(2)\text{, }\ldots\text{, }y(k)\]
\[\boldsymbol{\phi}(0)\text{, }\boldsymbol{\phi}(1)\text{, }\ldots\text{, }\boldsymbol{\phi}(k-1)\]
The problem is to estimate the \textbf{unkown} parameter
$\boldsymbol{\theta}$ using collected data above. Let the estimate of
$\boldsymbol{\theta}$ at the time instant $k$ be
$\hat{\boldsymbol{\theta}}\left(k\right)$ with all the collected data.
The estimation is conducted in the least-square sense, thus
\begin{align}
\hat{\boldsymbol{\theta}}\left(k\right)&=\arg\min_{\hat{\boldsymbol{\theta}}}\sum_{j=1}^{k}\frac{1}{2}\left(y(j)-\boldsymbol{\phi}\left(j-1\right)^T\hat{\boldsymbol{\theta}}\right)^2 \nonumber\\
&=\arg\min_{\hat{\boldsymbol{\theta}}}J\left(\hat{\boldsymbol{\theta}}\right) 
\end{align}
where
$J\left(\hat{\boldsymbol{\theta}}\right)=\sum_{j=1}^{k}\frac{1}{2}\left(y(j)-\boldsymbol{\phi}\left(j-1\right)^T\hat{\boldsymbol{\theta}}\right)^2$.
The optimal estimation is theofore obtained by solving
\begin{equation}
\frac{\partial J\left(\hat{\boldsymbol{\theta}}\right)}{\partial\hat{\boldsymbol{\theta}}}=0
\end{equation}
It results in
\begin{align}
\sum_{j=1}^{k}\left(y(j)-\boldsymbol{\phi}\left(j-1\right)^T\hat{\boldsymbol{\theta}}\right)\boldsymbol{\phi}\left(j-1\right)&=0
\end{align}
and
\begin{align}
\sum_{j=1}^{k}\boldsymbol{\phi}\left(j-1\right)y(j)&=\left(\sum_{j=1}^{k}\left(\boldsymbol{\phi}j-1\right)\boldsymbol{\phi}\left(j-1\right)^T\right)\hat{\boldsymbol{\theta}} \label{eq:bathLSE0}
\end{align} 
so
\begin{equation}
\hat{\boldsymbol{\theta}}\left(k\right) =\left(\sum_{j=1}^{k}\boldsymbol{\phi}\left(j-1\right)\boldsymbol{\phi}\left(j-1\right)^T \right)^\dagger\sum_{j=1}^{k}\boldsymbol{\phi}\left(j-1\right)y(j)\label{eq:bathLSE1}
\end{equation}
where "$\dagger$" denotes the Moore-Penrose pseudo inverse that could be resolved via the singular value decomposition. Let
$\boldsymbol{Y}\left(k\right)=\left[y(1),y(2),\ldots,y(k)\right]^T\in \mathbb{R}^{k\times 1}$
and
$\boldsymbol{\Phi}\left(k-1\right)=\left[\boldsymbol{\phi}\left(0\right),\boldsymbol{\phi}\left(1\right),\ldots,\boldsymbol{\phi}\left(k-1\right)\right] \in \mathbb{R}^{k\times n}$.
We will have
\begin{align}
J\left(\hat{\boldsymbol{\theta}}\right)&=\frac{1}{2}\Vert\boldsymbol{Y}\left(k\right)-\boldsymbol{\Phi}\left(k-1\right)^T\hat{\boldsymbol{\theta}}\left(k\right) \Vert_2^2 \label{eq:batchLSE_obj}
\end{align}
Hence, we also have
\begin{equation}
\hat{\boldsymbol{\theta}}\left(k\right) =\left(\boldsymbol{\Phi}\left(k-1\right)\boldsymbol{\Phi}\left(k-1\right)^T\right)^\dagger\boldsymbol{\Phi}\left(k-1\right)\boldsymbol{Y}\left(k\right) \label{eq:batchLSE}
\end{equation}
If we have collected sufficient data and the data has sufficient richness, 
$\boldsymbol{\Phi}\left(k-1\right)\boldsymbol{\Phi}\left(k-1\right)^T$
could be nonsingular. In this case, we need to calculate its inverse instead of psuedo
inverse. \hl{The result above is called \emph{\textbf{batch least square
estimation}}}. The computation complexity of the inverse in (\ref{eq:batchLSE}) is of $O\left(n^3\right)$, so the batch LSE is computationally expensive for high-dimensional cases.

%Note that if the least square error in (\ref{eq:batchLSE_obj})  is weighted by $\boldsymbol{W}\in\mathbb{R}^{k\times k}$, one has 
%\begin{align}
%J\left(\hat{\boldsymbol{\theta}}\right)=&\frac{1}{2}\left(\boldsymbol{Y}\left(k\right)-\boldsymbol{\Phi}\left(k-1\right)^T\hat{\boldsymbol{\theta}}\left(k\right) \right)^T\boldsymbol{W}\nonumber\\
%&\times\left(\boldsymbol{Y}\left(k\right)-\boldsymbol{\Phi}\left(k-1\right)^T\hat{\boldsymbol{\theta}}\left(k\right) \right) \label{eq:batchLSE_weightObj}
%\end{align}
%In this case, the batch LSE is rewritten as
%\begin{equation}
%\hat{\boldsymbol{\theta}}\left(k\right) =\left(\boldsymbol{\Phi}\left(k-1\right)\boldsymbol{W}\boldsymbol{\Phi}\left(k-1\right)^T\right)^\dagger\boldsymbol{\Phi}\left(k-1\right)\boldsymbol{W}\boldsymbol{Y}\left(k\right) \label{eq:batchLSE_weight}
%\end{equation}
%In the sequel, it is assumed that $\boldsymbol{W}$ is an identity matrix. 
%%-------8-------8-------8-------8-------8------- New Section -------8-------8-------8-------8-------8-------8-------
\section{Recursive least square estimation} \label{sec:RecurLSE}
The  \textbf{recursive least square estimation (RLSE)} is introduced to reduce the computational burden and storage requirement of batch LSE. It is an iterative implementation of batch LSE, which could be potentially applied online in a real-time fashion. An early report on recursive LSE can be found in \cite{Albert1965SIAM}. Assume that we have collected sufficient data
and the data has sufficient richness so that $\sum_{j=1}^{k}\boldsymbol{\phi}\left(j-1\right)\boldsymbol{\phi}\left(j-1\right)^T $ is invertible. 
Define
\begin{align}
\boldsymbol{F}\left(k\right)^{-1} &= \sum_{j=1}^{k}\boldsymbol{\phi}\left(j-1\right)\boldsymbol{\phi}\left(j-1\right)^T \nonumber\\
&=\boldsymbol{F}\left(k-1\right)^{-1}+\boldsymbol{\phi}\left(k-1\right)\boldsymbol{\phi}\left(k-1\right)^T \label{eq:Finv}
\end{align}
where $\boldsymbol{F}\left(k\right)$ is an inverse of $\sum_{j=1}^{k}\boldsymbol{\phi}\left(j-1\right)\boldsymbol{\phi}\left(j-1\right)^T $.
According to the bath least square estimation (\ref{eq:bathLSE1}), one has
\begin{equation}
\hat{\boldsymbol{\theta}}\left(k\right) =\boldsymbol{F}\left(k\right) \sum_{j=1}^{k}\boldsymbol{\phi}\left(j-1\right)y(j)
\end{equation}
For the RLSE, it is expected to have 
\begin{equation}
\hat{\boldsymbol{\theta}}\left(k\right) =\hat{\boldsymbol{\theta}}\left(k-1\right)+\boldsymbol{\varepsilon}(k)
\end{equation}
where $\boldsymbol{\varepsilon}(k)$ is a correction term, $\hat{\boldsymbol{\theta}}\left(k-1\right)$ is the estimation results at the last step, and $\hat{\boldsymbol{\theta}}\left(k\right) $ is the current estimation. 

The
\emph{\textbf{a-priori}} output prediction is defined with the state-of-the-art parameter estimation  $\hat{\boldsymbol{\theta}}\left(k-1\right)$  and latest measurement $\boldsymbol{\phi}^{T}\left(k-1\right)$, which is given as below.
\begin{equation}
\hat{y}\left({k\vert k-1}\right) = \boldsymbol{\phi}^{T}\left(k-1\right)\hat{\boldsymbol{\theta}}\left(k-1\right)
\end{equation}
So \emph{\textbf{a-priori}} output estimation error is
\begin{align}
\boldsymbol{e}\left({k\vert k-1}\right) &={y}\left(k\right)-\hat{y}\left({k\vert k-1}\right)\nonumber \\
&={y}\left(k\right)-\boldsymbol{\phi}^{T}\left(k-1\right)\hat{\boldsymbol{\theta}}\left(k-1\right)
\end{align}
According to (\ref{eq:bathLSE0}), one has
\begin{equation}
\boldsymbol{F}\left(k\right)^{-1}\hat{\boldsymbol{\theta}}\left(k\right) =\sum_{j=1}^{k}\boldsymbol{\phi}\left(j-1\right)y(j)
\end{equation}
In terms of (\ref{eq:Finv}), there exists
\begin{align}
\boldsymbol{F}\left(k\right)^{-1}\hat{\boldsymbol{\theta}}\left(k\right) 
=&\boldsymbol{F}\left(k-1\right)^{-1}\hat{\boldsymbol{\theta}}\left(k-1\right)+\boldsymbol{\phi}\left(k-1\right)y(k) \nonumber \\
=& \left(\boldsymbol{F}\left(k\right)^{-1}-\boldsymbol{\phi}\left(k-1\right)\boldsymbol{\phi}\left(k-1\right)^T\right)\nonumber \\
& \times\hat{\boldsymbol{\theta}}\left(k-1\right)+\boldsymbol{\phi}\left(k-1\right)y(k) \label{eq:}
\end{align}
Since $\hat{\boldsymbol{\theta}}\left(k\right) =\sum_{j=1}^{k}\boldsymbol{\phi}\left(j-1\right)y(j)$, one has
\begin{align}
\hat{\boldsymbol{\theta}}\left(k\right) 
=&\boldsymbol{F}\left(k\right) \left(\boldsymbol{F}\left(k\right)^{-1}-\boldsymbol{\phi}\left(k-1\right)\boldsymbol{\phi}\left(k-1\right)^T\right)\nonumber \\
& \times\hat{\boldsymbol{\theta}}\left(k-1\right)+\boldsymbol{F}\left(k\right)\boldsymbol{\phi}\left(k-1\right)y(k)\nonumber \\
&=\hat{\boldsymbol{\theta}}\left(k-1\right)-\boldsymbol{F}\left(k\right)\boldsymbol{\phi}\left(k-1\right)\boldsymbol{\phi}\left(k-1\right)^T\hat{\boldsymbol{\theta}}\left(k-1\right)\nonumber \\
&+\boldsymbol{F}\left(k\right)\boldsymbol{\phi}\left(k-1\right)y(k)\nonumber \\
&=\hat{\boldsymbol{\theta}}\left(k-1\right)+\boldsymbol{F}\left(k\right)\boldsymbol{\phi}\left(k-1\right)\nonumber \\
& \times\underbrace{\left(y(k)-\boldsymbol{\phi}\left(k-1\right)^T\hat{\boldsymbol{\theta}}\left(k-1\right)\right)}_{\boldsymbol{e}\left({k\vert k-1}\right) }
\end{align}
Eventually, one has
\begin{align}
\hat{\boldsymbol{\theta}}\left(k\right) 
&=\hat{\boldsymbol{\theta}}\left(k-1\right)+\underbrace{\boldsymbol{F}\left(k\right)\boldsymbol{\phi}\left(k-1\right)\boldsymbol{e}\left({k\vert k-1}\right)}_{\boldsymbol{\varepsilon}(k)}\label{eq:RLSE_update}
\end{align}
The equation (\ref{eq:RLSE_update}) is taken as the estimation update. Since
$\boldsymbol{F}\left(k\right)^{-1}  =\boldsymbol{F}\left(k-1\right)^{-1}+\boldsymbol{\phi}\left(k-1\right)\boldsymbol{\phi}\left(k-1\right)^T$,
the following update equation is obtained for
$\boldsymbol{F}\left(k\right)$ based on the \textbf{Sherman-Morrison
formula} in Lemma \ref{lem: Sherman-Morrison}.
\begin{equation*} \small
\boldsymbol{F}\left(k\right) =\boldsymbol{F}\left(k-1\right)-\frac{\boldsymbol{F}\left(k-1\right)\boldsymbol{\phi}\left(k-1\right)\boldsymbol{\phi}\left(k-1\right)^T\boldsymbol{F}\left(k-1\right)}{1+\boldsymbol{\phi}\left(k-1\right)^T\boldsymbol{F}\left(k-1\right)\boldsymbol{\phi}\left(k-1\right)}
\end{equation*}
In summary, the \emph{\textbf{RLSE}} algorithm is summarized as below.
\begin{equation}
\left\{
\begin{array}{rcl}
\hat{y}\left({k\vert k-1}\right) &=& \boldsymbol{\phi}^{T}\left(k-1\right)\hat{\boldsymbol{\theta}}\left(k-1\right)\\
\boldsymbol{e}\left({k\vert k-1}\right) &=& {y}\left(k\right)-\hat{y}\left({k\vert k-1}\right)\\
\boldsymbol{F}\left(k\right) &=& \boldsymbol{F}\left(k-1\right)-\frac{\boldsymbol{F}\left(k-1\right)\boldsymbol{\phi}\left(k-1\right)\boldsymbol{\phi}\left(k-1\right)^T\boldsymbol{F}\left(k-1\right)}{1+\boldsymbol{\phi}\left(k-1\right)^T\boldsymbol{F}\left(k-1\right)\boldsymbol{\phi}\left(k-1\right)}\\
\hat{\boldsymbol{\theta}}\left(k+1\right) 
&=&\hat{\boldsymbol{\theta}}\left(k\right)+\boldsymbol{F}\left(k\right)\boldsymbol{\phi}\left(k-1\right)\boldsymbol{e}\left({k\vert k-1}\right)
\end{array}
\right. \label{eq:aPriorRLSE}
\end{equation}
Initially, we choose
$\boldsymbol{F}\left(0\right)=\boldsymbol{F}\left(0\right)^T>0$ and a
guess on $\hat{\boldsymbol{\theta}}\left(0\right)$. The equation (\ref{eq:aPriorRLSE}) is also called  a-priori version of the recursive LSE. Another version is the posteriori recursive LSE that could be derived by following a similar process.

The trace of $\boldsymbol{F}\left(k\right)$ is given as below.
\begin{align}
tr\left[\boldsymbol{F}\left(k\right)\right] =& tr\left[\boldsymbol{F}\left(k-1\right)\right] \nonumber\\
&-\frac{\Vert\boldsymbol{F}\left(k-1\right)\boldsymbol{\phi}\left(k-1\right)\Vert_2}{\vert 1+\boldsymbol{\phi}\left(k-1\right)^T\boldsymbol{F}\left(k-1\right)\boldsymbol{\phi}\left(k-1\right) \vert}
\end{align}
The trace of $\boldsymbol{F}\left(k\right)$ will decrease, if $\boldsymbol{F}\left(k-1\right)\boldsymbol{\phi}\left(k-1\right)$ is not zero. Hence, \hl{the RLSE will eventually stop updating, due to the decrease of the
magnitude of $\boldsymbol{F}\left(k\right)$.}

\section{Recursive least square estimation with forgetting factors}\label{sec:ForgLSE}
To resolve the potential issue of RLS, we modify the cost function by
introducing the \textbf{forgetting factor} that discounts the old data,
so
\begin{equation}
J\left(\hat{\boldsymbol{\theta}}\right)=\min_{\hat{\boldsymbol{\theta}}}\frac{1}{2}\sum_{j=1}^{k}\lambda^{k-j}\left(y(j)-\boldsymbol{\phi}\left(j-1\right)^T\hat{\boldsymbol{\theta}}\right)^2 
\end{equation}
where $0<\lambda\leq1$ is a constant forgetting factor. With this cost function, the equation (\ref{eq:bathLSE0}) is rewritten as
\begin{align}
\sum_{j=1}^{k}\lambda^{k-j}\boldsymbol{\phi}\left(j-1\right)y(j)&=\left(\sum_{j=1}^{k}\lambda^{k-j}\left(\boldsymbol{\phi}j-1\right)\boldsymbol{\phi}\left(j-1\right)^T\right)\hat{\boldsymbol{\theta}} \label{eq:bathLSE_weight}
\end{align} 
If $\left(\sum_{j=1}^{k}\lambda^{k-j}\left(\boldsymbol{\phi}j-1\right)\boldsymbol{\phi}\left(j-1\right)^T\right)$ is invertible, $\boldsymbol{F}\left(k\right)$ is redefined as below.
\begin{align}
\boldsymbol{F}\left(k\right)^{-1} &= \sum_{j=1}^{k}\lambda^{k-j}\boldsymbol{\phi}\left(j-1\right)\boldsymbol{\phi}\left(j-1\right)^T \nonumber\\
&=\lambda\boldsymbol{F}\left(k-1\right)^{-1}+\boldsymbol{\phi}\left(k-1\right)\boldsymbol{\phi}\left(k-1\right)^T \label{eq:Finv_weight}
\end{align}
The \emph{\textbf{a-priori}} output prediction, estimation error, and estimation update equations are all the same as the case of the recursive LSE without forgetting factor. The only difference comes from the updates of the gain matrix $\boldsymbol{F}\left(k\right)$. According to (\ref{eq:Finv_weight}) and Lemma \ref{lem: Sherman-Morrison}, one has the following update equation for $\boldsymbol{F}\left(k\right)$.
\begin{align}
\boldsymbol{F}\left(k\right) &=\frac{\boldsymbol{F}\left(k-1\right)}{\lambda}+\frac{\boldsymbol{F}\left(k-1\right)\boldsymbol{\phi}\left(k-1\right)\boldsymbol{\phi}\left(k-1\right)^T\boldsymbol{F}\left(k-1\right)}{\lambda^2+\lambda\boldsymbol{\phi}\left(k-1\right)^T\boldsymbol{F}\left(k-1\right)\boldsymbol{\phi}\left(k-1\right)}\label{eq:Fweight_update}
\end{align}
In summary, the recursive LSE with a forgetting factor is given as below.
\begin{equation}
\left\{
\begin{array}{rcl}
\hat{y}\left({k\vert k-1}\right) &=& \boldsymbol{\phi}^{T}\left(k-1\right)\hat{\boldsymbol{\theta}}\left(k-1\right)\\
\boldsymbol{e}\left({k\vert k-1}\right) &=& {y}\left(k\right)-\hat{y}\left({k\vert k-1}\right)\\
\boldsymbol{F}\left(k\right) &=& \frac{1}{\lambda}\left[\boldsymbol{F}\left(k-1\right)-\frac{\boldsymbol{F}\left(k-1\right)\boldsymbol{\phi}\left(k-1\right)\boldsymbol{\phi}\left(k-1\right)^T\boldsymbol{F}\left(k-1\right)}{\lambda+\boldsymbol{\phi}\left(k-1\right)^T\boldsymbol{F}\left(k-1\right)\boldsymbol{\phi}\left(k-1\right)}\right]\\
\hat{\boldsymbol{\theta}}\left(k+1\right) 
&=&\hat{\boldsymbol{\theta}}\left(k\right)+\boldsymbol{F}\left(k\right)\boldsymbol{\phi}\left(k-1\right)\boldsymbol{e}\left({k\vert k-1}\right)
\end{array}
\right. \label{eq:aPriorRLSE_forgetting}
\end{equation}
\hl{The introduction of the forgetting factor $\lambda$ makes the recursive LSE possible for estimating time-varying parameters.}

%The derivation process is very similar to the case of RLS, so it is
%ignored. Readers of interest can refer to
%\emph{\textcolor{blue}{https://berkeley-me233.github.io/static/ME233_Sp16_L19_Least_Squares_Parameter_Estimation.pdf}}{HERE},
%\emph{\textcolor{blue}{https://www.it.uu.se/edu/course/homepage/systemid/vt12/Sysid_s2012_7.pdf}}}{HERE},
%or
%\emph{\textcolor{blue}{https://ocw.mit.edu/courses/electrical-engineering-and-computer-science/6-241j-dynamic-systems-and-control-spring-2011/readings/MIT6_241JS11_chap02.pdf}}{HERE}.
%
%\section{Application examples}
%\section{Conclusions}

\bibliographystyle{IEEEtran}
\bibliography{myRefs.bib}

\end{document}